# Time Perception in Virtual Reality: Effects of Emotional Valence and Stress Level

Kyriaki Syrigou[1†], Marina Stoforou[1], and Panagiotis Kourtesis[1-3†]
[1] Department of Psychology, The American College of Greece, Athens, Greece
[2] Department of Informatics & Telecommunications, National and Kapodistrian University of Athens, Athens, Greece
[3] Department of Psychology, The University of Edinburgh, Edinburgh, United Kingdom
[†] The authors contributed equally.
Corresponding author: Panagiotis Kourtesis; 6 Gravias Street, Athens, Greece, GR 15 342; tel: +30 210 600 9800, ext. 1488; Email: pkourtesis@acg.edu

## Abstract

Immersive virtual reality (VR) enables controlled tests of how emotional context distorts subjective time, yet findings remain mixed. Fifty-seven adults explored three five-minute VR environments (tranquil garden, neutral room, threatening sewer) while eye tracking recorded pupil diameter. Baseline stress was assessed before VR exposure. After each environment, participants estimated elapsed time and rated mood, calmness, and arousal. Repeated-measures ANCOVA (baseline stress as covariate) and linear mixed-effects models converged on a valence gradient: garden intervals were overestimated, sewer intervals underestimated, and neutral estimates were intermediate. Pupil diameter showed the complementary pattern (sewer > garden > neutral) and covaried with the magnitude of temporal distortion. Baseline stress did not predict time estimates and did not moderate valence effects. Results indicate that affectively distinct VR contexts reliably shift perceived time and elicit convergent physiological and self-report responses. These findings inform emotion-aware design of immersive experiences and digital-wellbeing safeguards that support safe, sustained immersion.

## Introduction

Time perception is elastic: identical chronometric intervals can feel markedly shorter or longer depending on emotional and attentional context, a phenomenon termed *emotional temporal distortion* [54]. Two complementary accounts explain these distortions. The Internal Clock Model (ICM) posits a pacemaker whose pulses pass through a switch to an accumulator and are compared against reference memory to yield duration judgements [38, 65, 94, 95]. Converging evidence suggests that arousal increases pacemaker rate and can lengthen perceived duration, whereas lower arousal can slow this process [28, 40, 73]. Because pacemaker dynamics alone cannot explain why absorbing experiences often feel brief, the Attentional Gate Model (AGM) proposes a capacity-limited gate: attending to time widens the gate and admits more pulses, whereas attention captured by non-temporal content narrows it [10].

Together, these mechanisms imply that pleasant, engaging activity often compresses duration by reducing temporal attention, whereas threat or intense negative affect can promote dilation via heightened arousal and vigilant monitoring [6, 23, 27]. Empirically, however, time–emotion effects remain inconsistent, likely because many paradigms rely on weak affect inductions, "neutral" baselines that vary in engagement, and limited control over attentional demand [18, 82]. Immersive virtual reality (VR) offers a useful testbed by sustaining naturalistic affect and engagement over minutes while preserving experimental control; importantly, early evidence suggests that high immersion may prioritise attentional capture over arousal-driven timing changes. For example, [99] reported increases in duration-judgement ratios with exposure length that were only weakly related to self-reported flow, raising the possibility that attentional allocation is a primary driver of temporal distortion in VR.

Behavioural findings converge with this interpretation: dual-task and working-memory loads, immersive videogames, and flow states typically divert attention from time and yield systematic underestimation [14, 15, 29, 96], whereas contexts that enforce vigilant monitoring—such as sustained pain, threat of shock, or monotonous boredom—tend to produce robust overestimation [6, 24, 25, 107]. Critically, attentional capture can override arousal-linked influences: highly arousing yet absorbing experiences can feel fleeting, while low-arousal boredom can make time drag [102]. Taken together, these patterns indicate that subjective time reflects not only internal-clock dynamics but also flexible attentional deployment as a proximate determinant of perceived duration.

**Emotional Valence, Trait Stresss, and Temporal Distortion**

Emotional valence can bias perceived duration by shaping both affective arousal and what people allocate attention to during an interval—two routes that map naturally onto arousal-linked timing and attention-to-time accounts [9, 19, 31, 57]. Early work framed a relatively simple pattern in which pleasant states compress time and unpleasant states dilate it [27, 36]. Many studies still support this "classic" direction, with negative affect often linked to longer prospective estimates and salience-related processing [6, 35, 62, 107]. Yet valence effects are reliably bidirectional: engaging negative material (e.g., absorbing films or recalled sadness) can be associated with faster subjective time, and anger- or pain-related cues sometimes yield underestimation [5, 8, 34, 110].

A parsimonious synthesis is that the *direction* of temporal distortion often tracks engagement: boredom and low novelty make time drag, whereas challenge and "flow" can make time pass quickly even under negative affect [22, 44, 105]. Within the ICM–AGM framing, valence can shift arousal, while engagement regulates attention to time—so the observable bias reflects which influence dominates in a given setting [10, 29]. Inconsistencies are further amplified when affect induction is unverified, stimulus properties are mismatched, or exposures are too brief to stabilise affective and attentional states [54, 64, 93].

Trait stress is also valuable to examine alongside valence as a complementary factor, because it can alter arousal and attentional deployment in ways that may amplify—or counteract—valence effects rather than producing a uniform main effect. Acute stress can either compress time (when action-focused responding monopolises attention) or dilate it (when vigilance and temporal monitoring increase), depending on task demands [42]. Experimentally, acute stressors often lengthen produced/reproduced intervals, consistent with heightened arousal and monitoring [106, 108, 109], whereas chronic stress has been linked to broader alterations in temporal sensitivity and cognitive control [32, 66]. Stress-related shifts can also be tracked physiologically: pupil dilation indexes arousal and attentional effort [63] and has been connected to anxiety/stress processes relevant to timing biases [1, 60].

Empirically, effects vary by context and measure: during COVID-19 lockdowns, higher perceived stress and isolation were associated with a slowed passage of time [26], and clinical groups marked by heightened stress reactivity frequently overestimate durations [52, 70, 81, 98], although intense hypervigilance can also yield underestimation when threat processing consumes resources. Because stress and negative valence often co-occur, they may interact in additive or synergistic ways: stress can magnify distortions to unpleasant cues [6], including in social anxiety and related contexts [43, 45, 111].

At the same time, arousal- and attention-related influences can oppose one another—highly arousing unpleasant contexts may maximise dilation, whereas low-arousal unpleasant material or highly engaging pleasant contexts can weaken or reverse the effect [2, 55]. A key limitation is that much of the stress and/or valence related evidence comes from sub-second to few-second intervals, which may miss the slower evolution of affect and attentional engagement [78]. Extending these questions to multi-minute exposures is therefore important, particularly given field-inspired findings that stress effects depend on concurrent monitoring demand [69]. Finally, trait stress/anxiety biases emotional processing [30, 59], , motivating tests of whether stress moderates valence-related timing effects under more ecological conditions.

## Virtual Reality as an Ecologically Valid Tool for Studying Emotion and Time Perception

Many inconsistencies in the time–emotion literature stem from the stimuli themselves: static pictures or brief clips rarely combine strong valence, sustained arousal, and meaningful engagement within a single design. Given that multi-minute exposure is often required for affective and attentional states to stabilise, many laboratory paradigms conflate emotional valence with differences in engagement and attentional demand, making it difficult to adjudicate ICM- versus AGM-consistent mechanisms.

In contrast, immersive VR offers an ecologically valid combination—naturalistic sensorimotor contingencies with experimental control—that may help with reducing the common trade-off between realism and mechanistic inference in emotion–time research. VR environments allow exposure duration and interactivity to be systematically controlled [48, 61, 89], while behavioural and physiological responses often resemble real-world responding [50, 87]. More broadly, VR scenes can reliably modulate cognition, mood, and autonomic activity, particularly when presence and agency are high [50, 80, 86, 88]. By varying emotional context while shaping attentional capture, VR helps adjudicate arousal- versus attention-based accounts of temporal distortion. Consistent with this potential, prior work suggests that relaxing VR contexts can make sessions feel shorter, whereas suspenseful or game-like VR can make longer intervals feel surprisingly brief [79, 83].

Finally, by combining self-reported affect with continuous indices such as pupil size (and related behavioural traces), which modern VR headsets facilitate, it allows an examination of whether timing biases track arousal, attentional engagement, or their combination. VR therefore enables a stringent test of whether temporal distortion in emotional contexts is better explained by arousal-linked timing changes, attentional engagement, or their combination—questions addressed in the present study.

## Aims and Hypotheses of the Study

Evidence remains mixed on how emotional valence and stress—separately and jointly—shape perceived time, mood, and physiological arousal. This likely reflects common limitations: many paradigms use brief/static stimuli, rely on self-report alone, do not account for VR-relevant scene confounds (e.g., sensory complexity, spatial openness, luminance/contrast), and often operationalise stress categorically, masking continuous individual differences and interactions. Prior work nonetheless suggests that emotional context biases subjective time and affect, that engagement can amplify or reverse arousal-linked timing effects, and that stress-related individual differences may shape reactivity—although directions vary across tasks and measures.

In response to these knowledge gaps and limitations, the present study used immersive VR to examine arousal- and attention-based accounts of emotional temporal distortion by combining multi-minute emotion induction with self-report (mood, calmness, arousal) and a physiological index of arousal/attentional engagement (pupil diameter). Participants explored calming, neutral, and stressful environments and provided verbal time estimates after each exposure. We tested (i) whether these contexts produce a reliable valence gradient across behavioural, subjective, and pupillometric outcomes, and (ii) whether baseline perceived stress—treated as a continuous covariate—predicts overall shifts and/or moderates valence-related effects. Accordingly, the study examined the following non-directional hypotheses:

***H1 (Valence main effects).*** Perceived duration, mood, calmness, arousal, and pupil diameter will differ across the calming, neutral, and stressful VR environments.

***H2 (Baseline stress main effects).*** Baseline perceived stress will predict overall differences in perceived duration, mood, calmness, arousal, and pupil diameter, independent of environment.

***H3 (Valence × baseline stress).*** Baseline perceived stress will moderate valence-related differences in perceived duration, mood, calmness, arousal, and pupil diameter.

## Methods

A one-minute video abstract offers a narrated walkthrough of the entire protocol—teleportation, sequential exploration of the Calming garden, Neutral studio, and Stressful sewer, distractor task, plus the in-headset prompts for the time-estimation and mood scales. The video abstract can be watched at Abstract. All non-proprietary study materials—including scene screenshots, layout figures, and all audio stimuli (sound-effects and music tracks)—are archived on the Open Science Framework (OSF) at OSF.

### Participants

The study was approved by the Ethics Committee of the American College of Greece and complied with the Declaration of Helsinki and the American Psychological Association's Ethical Principles of Psychologists and Code of Conduct. Volunteers were recruited through social-media advertisements, institutional e-mail lists, campus posters, and the institution's participant pool. Inclusion criteria required individuals to be between 18 and 45 years of age, to have normal or corrected-to-normal vision, and to report no history of neurological disorders such as multiple sclerosis or Parkinson's disease, psychiatric disorders such as major depression or generalized anxiety disorder, cardiovascular disease including hypertension or stroke, neurodevelopmental conditions such as autism-spectrum disorder or attention-deficit/hyperactivity disorder, neurodegenerative diseases such as dementia or mild cognitive impairment, diagnosed learning difficulties such as dyslexia or dyscalculia, or current alcohol or drug abuse. All testing was conducted in the PsyNet Lab of the Department of Psychology. After providing written informed consent, 57 young adults (30 women, 27 men) completed the study. Their ages ranged from 18 to 39 years (M = 25.07, SD = 4.34).

### Material

#### *Hardware*

This study employed a fully immersive VR setup designed to minimize cybersickness and deliver high-fidelity experiences. At its core was a high-performance desktop PC (Intel64 Family 6 Model 167 Stepping 1 processor at ~3.5 GHz, 32 GB RAM) running SteamVR (Valve Corporation, Bellevue, WA, USA) under Windows. Visual rendering, head tracking, and binocular eye-tracking data were handled by a Varjo Aero headset (Varjo Technologies Oy, Helsinki, Finland), which features dual mini-LED panels at 2,880 × 2,720 pixels per eye, a 115° horizontal field of view, 90 Hz refresh rate, custom low-distortion optics, and an embedded Tobii eye-tracker with 200 Hz refresh rate. Eye-tracking quality was verified via standard calibration prior to each condition. Pupil samples associated with blinks/loss of tracking were removed, and pupil diameter was aggregated per condition after preprocessing. Positional tracking used two HTC Vive Lighthouse base stations (HTC Corporation, New Taipei City, Taiwan) arranged diagonally to cover a large play area with sub-millimeter precision. Auditory immersion was provided by Sony noise-canceling headphones (Sony Corporation, Tokyo, Japan), and interaction and locomotion were realized through ergonomically designed HTC Vive controllers (HTC Corporation, New Taipei City, Taiwan), complete with haptic feedback, grip buttons, trackpads, and triggers.

#### *Cybersickness in VR Questionnaire (CSQ-VR)*

To quantify any cybersickness symptoms induced by the VR exposure, we administered the Cybersickness in VR Questionnaire (CSQ-VR) both immediately before and immediately after all virtual-reality sessions [51, 72]. The CSQ-VR consists of six items rated on a 7-point Likert scale (1 = Absent feeling; 7 = Extreme feeling) and probes three symptom domains: nausea, vestibular disturbance, and oculomotor discomfort. For example, participants responded to prompts such as "Do you experience dizziness (e.g., light-headedness or a spinning sensation)?" by marking the intensity of their current experience. The CSQ-VR exhibits high internal consistency (α = .865), making this measure more sensitive to VR-induced performance decrements than earlier cybersickness scales [51, 72].

#### *Perceived Stress Scale (PSS-10)*

Baseline perceived stress was assessed using the 10-item Perceived Stress Scale (PSS-10; Cohen et al., 1983), which captures both feelings of helplessness and perceived self-efficacy over the preceding month. Six negatively phrased items (e.g., "In the last month, how often have you felt nervous and stressed?") index perceived helplessness, while four positively phrased items (e.g., "In the last month, how often have you felt confident about your ability to handle your personal problems?") assess coping efficacy. Respondents rate each item on a five-point Likert scale ranging from 0 ("never") to 4 ("always"), with total scores reflecting overall perceived stress. The PSS-10 was chosen for its strong psychometric properties, such as internal consistency (Cronbach's α = .67–.91;Denovan et al., 2017; Taylor, 2015), concurrent validity with anxiety and depression measures [58, 67], and convergent validity with objective stressors [4].

***Trail Making Test in Virtual Reality (TMT-VR)***

The TMT-VR is a head-controlled, immersive adaptation of the Trail Making Test that preserves the core cognitive demands (visual scanning, processing speed, and set-switching) while increasing task ecological validity through embodied, 3D interaction [37, 39, 49]. In a 360° virtual space, twenty-five targets are distributed across depth and plane (see Figure 1). Participants select targets sequentially using a fixed central reticle controlled by head orientation with a brief dwell-time confirmation. Task A requires selecting numbers in ascending order (1–25), whereas Task B alternates numbers and letters (1–A–2–B–…–13). Target locations are re-randomised at each administration to limit spatial learning, and the system automatically records completion time and errors.

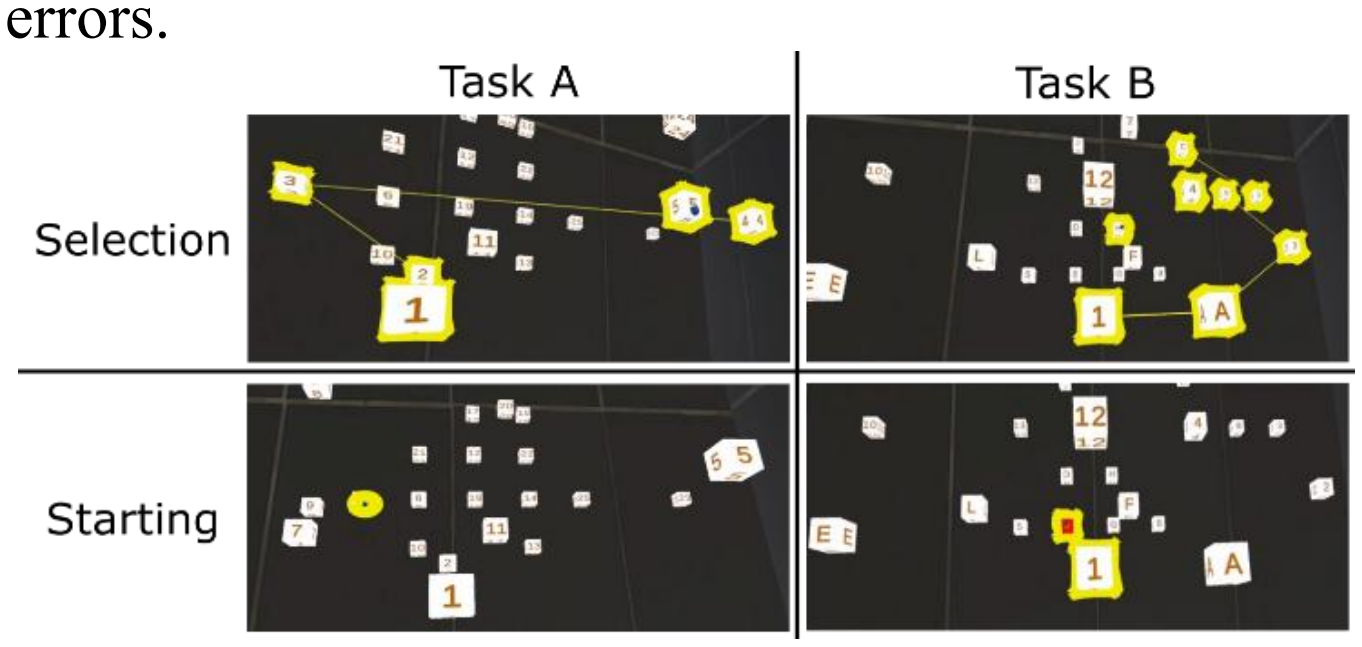

**Figure 1.** TMT-VR. *Task A (Left), Task B (Right), Selection of Targets (Top), and Starting Position*

***Emotional Valence in VR***

The experiment probed temporal judgments under three affective virtual-reality conditions (i.e., Calming, Neutral, and Stressful-Thrilling), while keeping exposure length, viewing pace, and locomotor effort comparable. Navigation relied exclusively on teleportation: each environment contained a fixed path of 48 teleport points that appeared sequentially; after a participant jumped to one point, the next became visible 5 seconds later, standardising the minimum dwell time at each point. In the Calming condition, participants wandered through a tranquil Japanese garden rendered with blossom trees, small ponds, grass, drifting fireflies, and a wooden temple; continuous shakuhachi flute and soft water sounds reinforced the calming ambience (see Figure 2). The Neutral condition placed them in an undecorated rectangular room with natural daylight and muted wall colours; no ambient sound was presented, and the scene was intentionally devoid of features likely to trigger pleasant or unpleasant affect (see Figure 3). The Stressful-Thrilling condition immersed participants in a dim underground sewer comprising narrow brick corridors scattered with smashed bottles and refuse; the hand controller doubled as a flashlight, and spatialised audio delivered dripping water, metallic clangs, and intermittent rat scurries to induce tension (see Figure 4). Apart from their emotional content and soundtrack, the three worlds were matched for geometric complexity, texture resolution, teleport count, and total exposure duration, allowing observed differences in perceived time, mood, or physiology to be attributed to emotional valence rather than to sensory load or locomotion.

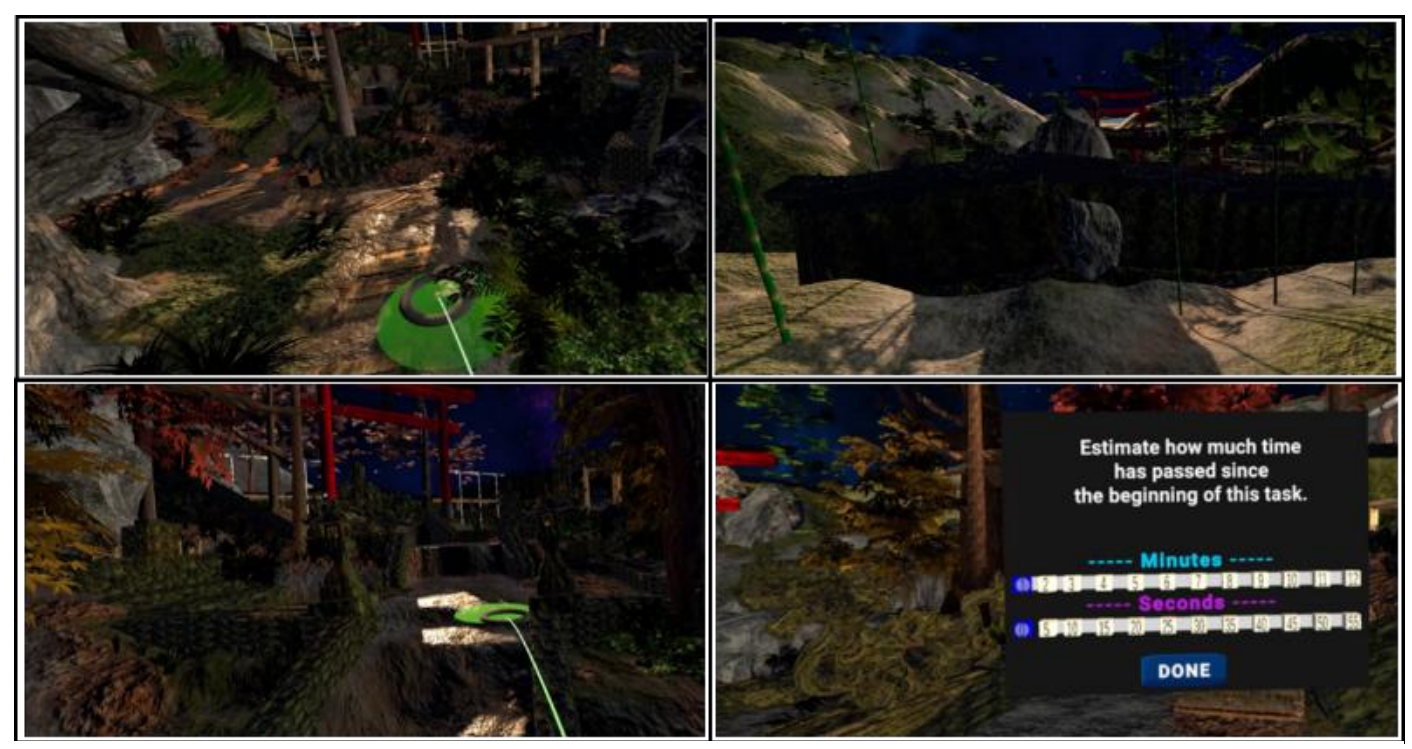

**Figure 2.** Japanese Garden Environment. *The tranquil Japanese garden during exploration via teleportation, as well as the in-world time estimation prompt are displayed.*

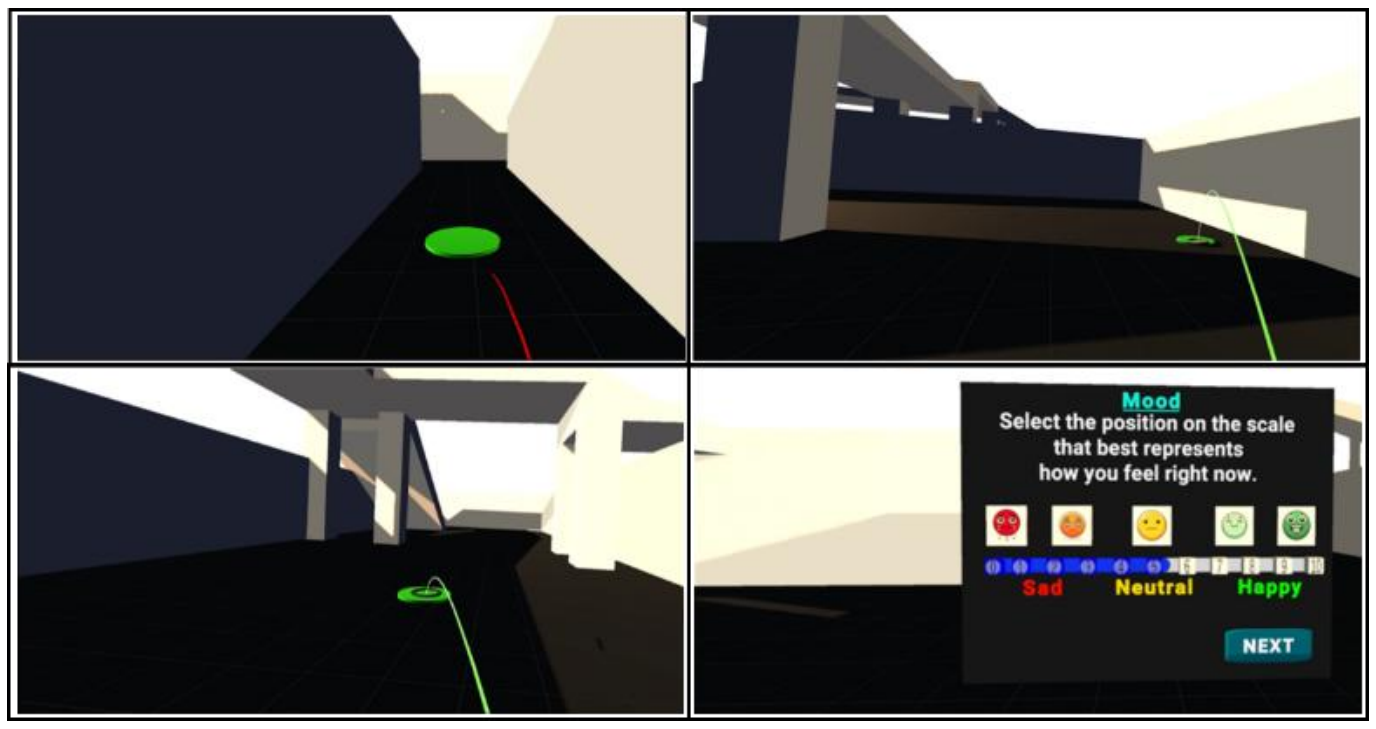

**Figure 3** Neutral Environment. *A plain, minimally textured area with natural lighting and no audio cues, navigated via the same teleportation sequence, as well as a mood-rating scale presented within this environment.*

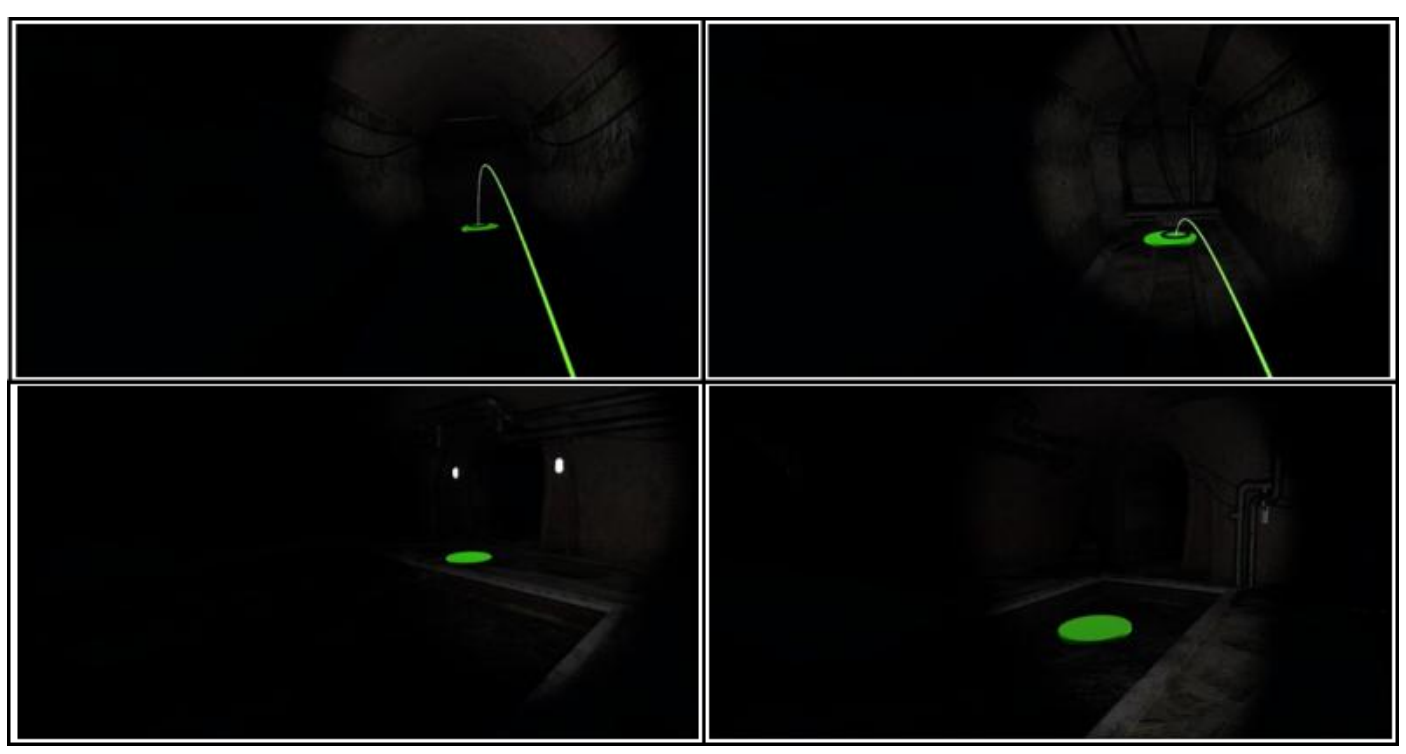

**Figure 4.** Stressful Underground Sewer Environment. *Dark tunnel segments, cluttered with debris and accompanied by ambient rat and drip sounds, navigated by head-based teleportation while wielding the controller as a flashlight.*

### *Verbal Estimation Task (VET)*

To measure participants' subjective perception of elapsed time, we employed a Verbal Estimation Task [103]. Verbal estimation was selected to minimise disruption of immersion during multi-minute exposures, since full-length reproduction would substantially increase testing time and fatigue and interval bisection is typically better suited to short, trial-based durations. Immediately after each environment of a different emotional valence, a prompt appeared asking, "Estimate how much time has passed since the beginning of this task." Participants then provided an estimate in minutes (1–12) and seconds (0–55) [97]. By embedding the VET directly into the virtual environment, users remained fully immersed, thereby maintaining presence while capturing their perceived duration (see Figure 5). Verbal estimates were converted to seconds and compared with the true exposure duration for each scene. The primary outcome was computed via a subtraction (i.e., *Time-Passage Difference = estimated duration - actual duration*), where positive values indicate overestimation and negative values indicate underestimation.

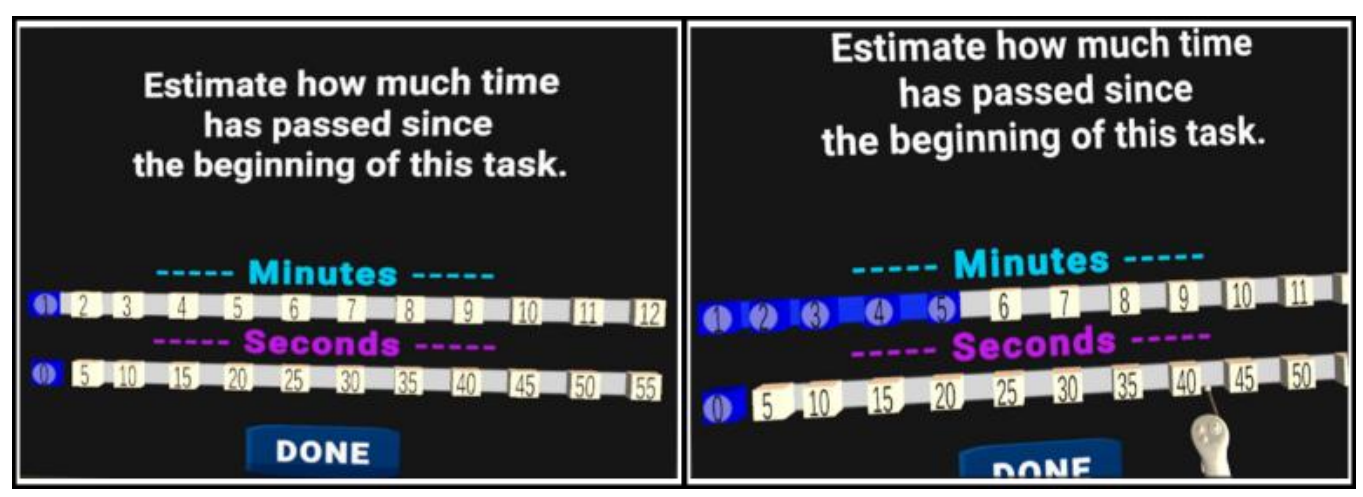

**Figure 5.** Time Estimation Task in VR. *Participants had to select one cube on the top row for minutes (1–12) and one on the bottom row for seconds (0–55), then press "Done" to record their estimate of how much time has elapsed.*

### *Visual Analog Mood Scales (VAMS)*

The three Visual Analog Mood Scales [90] were administered after each VET to assess current affective state. Each scale consisted of a 10-cm horizontal line anchored at one end with a descriptive label and ideogram—for mood (Sad–Happy), calmness (Stressed–Calm), and arousal (Sleepy–Energetic) [3]. Participants indicated their feeling, and scores were recorded. This simple, rapid format provided a sensitive check on each participant's emotional response without disrupting the VR experience (see Figure 6).

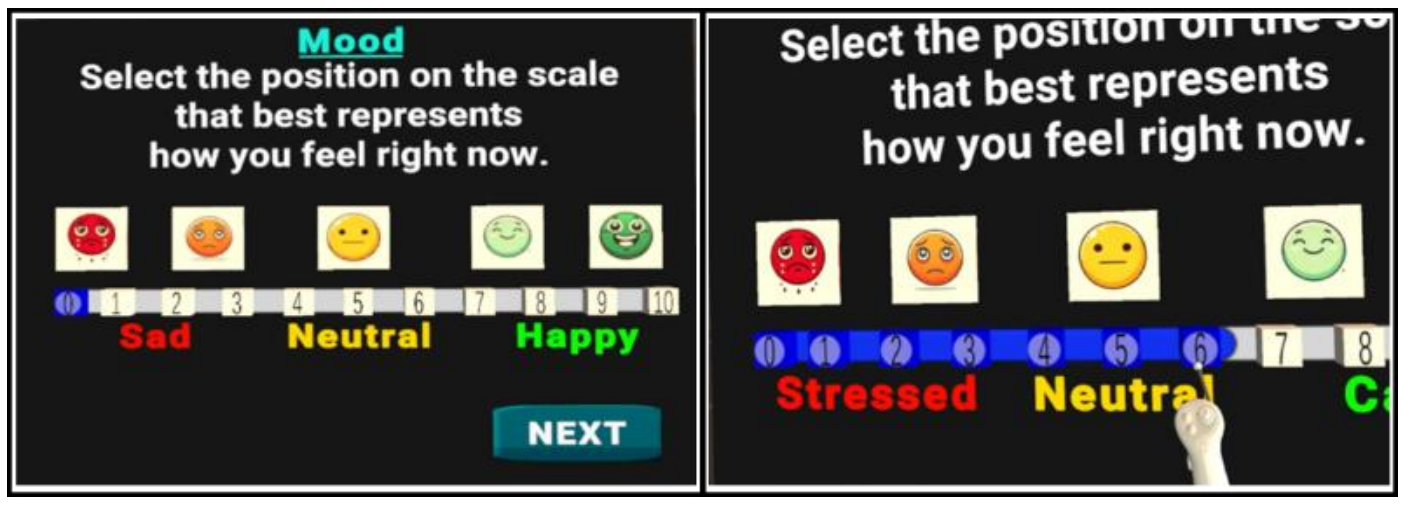

**Figure 6.** Visual Analog Mood Scales in VR. *The “Sad–Happy” and “Stressed–Calm” scales are depicted respectively, each annotated with coloured faces and numbered cubes. Participants had to select one cube to indicate their response.*

## Procedure

### *Pre-VR Exposure Phase*

On arrival, each volunteer signed an informed-consent form, was reminded of the right to withdraw at any time, and provided basic demographic information. To document baseline cybersickness symptoms, the CSQ-VR was administered. The PSS-10 was also administered to evaluate baseline perceived stress. The experimenter then introduced the virtual-reality hardware: head-mounted display, optical-tracking sensors, and hand controller.

### *VR Exposure Phase*

The exposure started with a short practice scene where participants rehearsed the required head-based interaction for the TMT-VR and were told that during every forthcoming virtual environment, they would have to judge how much time had passed; this prospective-timing instruction completed the preparation stage. Without removing the headset, each participant experienced the three emotional valence environments: Calming, Neutral, and Stressful. The order of the three emotional valence environments was in a counter-balanced order across participants. Scene order was counterbalanced to balance serial position using a 3-sequence Latin-square scheme (Calming–Neutral–Stressful; Neutral–Stressful–Calming; Stressful–Calming–Neutral). Participants were assigned to sequences in rotation as they enrolled. To further reduce carryover, a distractor task (TMT-VR A/B) separated successive environments.

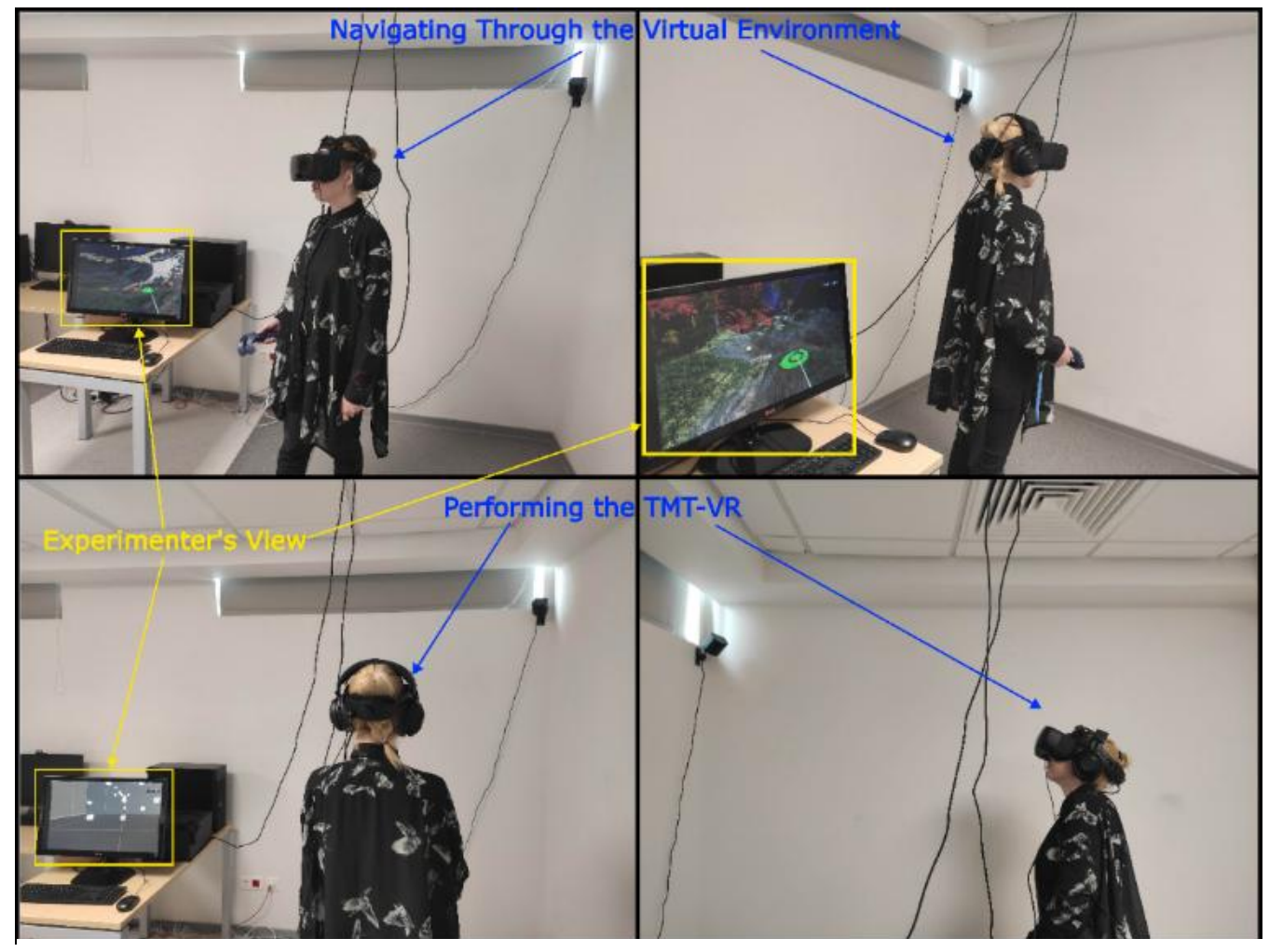

**Figure 7.** Experimental Setup. *The participant stands in the laboratory wearing the Varjo Aero headset (with embedded Tobii eye-tracker) and noise-cancelling headphones, using an HTC Vive controller to teleport through the virtual environment via green waypoint markers. The experimenter monitors the participant’s real-time first-person view on a desktop display to track navigation, task performance, and ensure protocol adherence.*

In each environment, the participant had to go through a predefined path to explore the environment through a brisk “walk” (see Figure 7). Every environment lasted approximately five minutes, which has been seen as adequate duration for investigating time passage perception in VR [68, 99], and was followed immediately by three tasks. First, the participant completed a VET, entering the number of minutes and seconds they believed had elapsed. Second, current affective state was recorded on three VAMS that indexed mood, perceived stress, and arousal. Third, to minimise carry-over of emotional state, the TMT-VR (Parts A and B) was presented during the interval that separated one environment from the next. The entire cycle (emotional valence environment and distractor task completion) was repeated until all three environments had been completed.

***Post-VR Exposure Phase***

After the headset was removed, the CSQ-VR was re-administered to record any cybersickness that might have developed during immersion, and participants were asked to confirm that they had followed the instructions seriously. A verbal debriefing then explained the study's aims, reiterated anonymity and confidentiality, and thanked them for their participation. The full procedure—from consent signing to debriefing—took approximately thirty minutes for each participant.

**Manipulation checks and confound monitoring**

Cybersickness was assessed with the CSQ-VR immediately before and after the VR session to evaluate whether discomfort could confound affective responding or time judgements across conditions [47, 51]. As a data-quality safeguard, unusually high post-exposure CSQ-VR scores were flagged for sensitivity analyses and, if severe, for exclusion to minimise discomfort-related confounding of mood and timing outcomes. Emotional-valence induction was evaluated using convergent indices: self-report ratings on the VAMS (VAMS mood, calmness, arousal; Athanasou, 2019) collected after each scene, and continuous pupillometry (pupil diameter) recorded during exposure. Pupil diameter was treated as a physiological correlate of arousal (Bradley et al., 2008; Strauch et al., 2022; Wang et al., 2018) and attentional engagement [13, 91, 100], interpreted as complementary to the self-reported measures.

**Statistical Analyses**

All analyses were performed in R 4.4.0 [77] within the RStudio 2024.03 environment [75]. Raw data were screened for entry errors and extreme outliers ($|z| > 3.29$). Univariate normality was assessed with the Shapiro–Wilk test; variables that departed significantly from normality were normalised with bestNormalize 1.9.1 [74]. All dependent measures were then z-standardised so that effect sizes are expressed in pooled SD units. Descriptive statistics (mean, SD, range) were computed for every outcome and visualised with ggplot2 3.5.0 [104]. Power and sensitivity analyses were conducted using pwr 1.3-0 (ANCOVA; Champely, 2020), as well as lme4 1.1-35.5 [7] and lmerTest 3.1-3 [53] packages for model fitting and data simulation (LMM sensitivity).

An a priori power analysis for the repeated-measures ANCOVA within-subject effect of Emotional Valence (three levels; $\alpha = .05$; $1 - \beta = .80$; medium effect $f = 0.30$) indicated a minimum required sample size of $N = 38$ participants. The final sample ($N = 57$) exceeded this threshold. Because closed-form a priori power formulas are not available for the planned mixed-effects moderation model, we additionally conducted a design-based simulation sensitivity analysis for the Valence × baseline stress interaction under the planned LMM structure (random intercept for participants; model comparison for interaction term; 2,000 simulations; standardized outcome and baseline stress). Estimated power was 0.198 for a standardised Valence × baseline stress fixed-effect coefficient of $\beta = 0.05$, 0.601 for $\beta = 0.10$, 0.935 for $\beta = 0.15$ (small-to-moderate), 0.996 for $\beta = 0.20$, and 1.000 for $\beta = 0.30$, indicating adequate power for small-to-moderate or larger moderation effects but limited sensitivity to small interactions.

Inferential analyses reflected the repeated-measures design. For each dependent variable (Time-Passage Difference, pupil diameter, VAMS Mood, VAMS Calmness, and VAMS Energy), we fitted a repeated-measures ANCOVA with Emotional Valence (Calming, Neutral, Stressful) as a within-subject factor and baseline perceived stress (PSS-10) as a continuous covariate, including the Valence × Stress interaction to test moderation by baseline stress. Repeated-measures ANCOVAs were estimated using afex package 0.22 [85]. When Mauchly's test indicated a violation of sphericity, Greenhouse–Geisser corrections were applied. Significant main effects or interactions were followed by pairwise contrasts obtained with emmeans package 1.10 [56]; p-values were adjusted with the Holm method to control family-wise error. Effect sizes are reported as partial eta-squared ($\eta p^2$) for ANCOVA analyses and Cohen's d for pairwise comparisons.

As a robustness check, we additionally fitted linear mixed-effects models (LMMs) for each outcome using random intercepts for participants to account for within-subject dependence. The LMMs were run with lme4 1.1-35.5 [7] and lmerTest 3.1-3 [53] packages. Fixed effects mirrored the ANCOVA specification (Valence, PSS, and Valence × PSS). Inference on fixed effects used Type III tests, and post hoc contrasts for Valence were computed via *emmeans* with Holm adjustment. Lastly, the validation of the emotional-valence manipulation and absence of cybersickness-related confounding were evaluated by testing the effect of Valence on VAMS outcomes and pupil diameter, and by confirming low cybersickness scores.

# Results

The scores on the CSQ-VR remained low throughout (absent to mild symptoms); no participant endorsed moderate or intense symptoms either before or after VR exposure, confirming that cybersickness did not confound our findings. Table 1 displays the descriptive statistics of the targeted variables. On average, participants overestimated the five-minute intervals. Self-reported mood was generally positive, while calmness ratings fell in the mid-range. Energy levels (i.e., arousal) were also moderate. Finally, pupil size (i.e., diameter) across conditions displayed adequate variability

Notably, emotional valence (see Table 2) drove pronounced differences across conditions. Calming scenes yielded the greatest overestimation of elapsed time, highest mood and calmness ratings, and intermediate pupil dilations. Neutral scenes produced moderate overestimations, mid-range mood and calmness, and the smallest pupil responses. Stressful scenes reversed this pattern, where participants underestimated time reported the lowest mood and calmness, showed elevated energy, and exhibited the largest pupil dilations. These descriptive trends hint the significant main effects of valence reported below.

Importantly, the valence manipulation was supported by convergent subjective and physiological responses. Across conditions, VAMS ratings shifted systematically: calming scenes increased mood and calmness, whereas stressful scenes decreased mood and calmness and increased energy (arousal). In parallel, pupil diameter differed robustly by condition, showing the largest dilation in the stressful scene, intermediate dilation in the calming scene, and the smallest response in the neutral scene, consistent with greater autonomic arousal/attentional engagement under emotionally salient contexts

**Effects on Mood**

A repeated-measures ANCOVA showed a significant main effect of valence on mood, $F(1.67, 91.67) = 21.25$, $p < .001$, $\eta p^2 = .28$. Neither the covariate, $F(1, 55) = 0.64$, $p = .427$, $\eta p^2 = .01$, nor the Valence × PSS interaction, $F(1.67, 91.67) = 0.04$, $p = .940$, $\eta p^2 < .001$, was significant. Holm-adjusted pairwise comparisons (Figure 8) indicated that mood was higher after the Calming than the Neutral condition, $t(55) = 2.36$, $p = .022$, $d = 0.32$; markedly higher after Calming than Stressful, $t(55) = 5.31$, $p < .001$, $d = 0.71$; and higher in Neutral than Stressful, $t(55) = 4.74$, $p < .001$, $d = 0.63$.

**Table 1. Descriptive Statistics**

| | Mean | SD | Minimum | Maximum |
|---|---|---|---|---|
| Time Passage Difference | 31.28 | 121.14 | -292.28 | 321.03 |
| Mood | 6.15 | 1.92 | 1 | 10 |
| Calmness | 5.62 | 2.22 | 0 | 10 |
| Energy | 6.09 | 1.85 | 0 | 10 |
| Pupil Size | 5.30 | 1.09 | 3.04 | 8.16 |

*Note.* Time Passage Difference measured in seconds. Pupil Size measured in millimetres.

**Table 2. Descriptive Statistics by Emotional Valence Condition**

| | Emotional Valence | Mean | SD | Minimum | Maximum |
|---|---|---|---|---|---|
| Time Passage Difference | Calming | 75.69 | 108.568 | -250.06 | 269.45 |
| | Neutral | 29.00 | 121.332 | -292.28 | 321.03 |
| | Stressful | -10.74 | 119.287 | -259.18 | 269.73 |
| Mood | Calming | 6.89 | 1.589 | 3 | 10 |
| | Neutral | 6.37 | 1.611 | 3 | 10 |
| | Stressful | 5.18 | 2.131 | 1 | 10 |
| Calmness | Calming | 6.63 | 1.848 | 3 | 10 |
| | Neutral | 5.91 | 2.003 | 0 | 10 |
| | Stressful | 4.32 | 2.148 | 0 | 9 |
| Energy | Calming | 5.82 | 1.794 | 1 | 10 |
| | Neutral | 5.91 | 1.714 | 2 | 10 |
| | Stressful | 6.54 | 1.983 | 0 | 10 |
| Pupil Size | Calming | 5.52 | 0.884 | 3.64 | 7.42 |
| | Neutral | 4.39 | 0.717 | 3.04 | 6.17 |
| | Stressful | 5.93 | 0.945 | 4.21 | 8.16 |

*Note.* Time Passage Difference measured in seconds. Pupil Size measured in millimetres

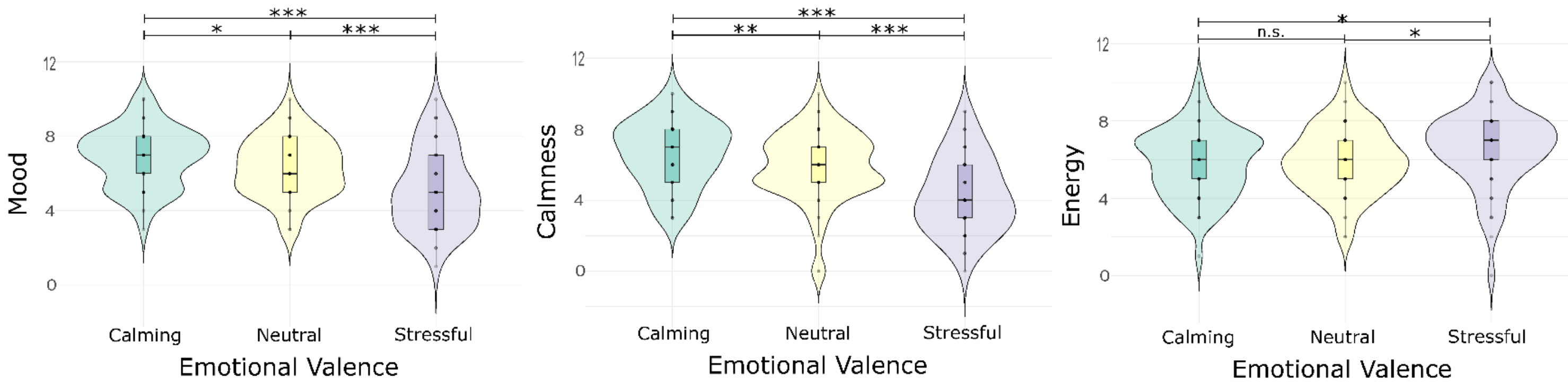

**Figure 8.** Emotional Valence Effects on Mood, Calmness, and Energy. ** p <.05; ** p <.01; *** p <.001; n.s.: non-significant.*

As a robustness check, the linear mixed-effects model with random intercepts converged on the same pattern, showing a strong valence effect, $\chi^2(2) = 44.04$, $p < .001$, with no main effect of PSS, $\chi^2(1) = 0.66$, $p = .415$, and no Valence × PSS interaction, $\chi^2(2) = 0.08$, $p = .960$. Overall, these results support H1: emotional valence systematically increased mood in calming settings and reduced it under stress, while H2 and H3 were not supported, as baseline stress neither shifted mood overall nor moderated valence effects.

**Effects on Calmness**

Perceived calmness varied significantly by emotional valence in the repeated-measures ANCOVA, $F(1.69, 93.22) = 26.29$, $p < .001$, $\eta p^2 = .32$. Neither the covariate, $F(1, 55) = 0.47$, $p = .497$, $\eta p^2 = .01$, nor the Valence × PSS interaction was significant, $F(1.69, 93.22) = 0.26$, $p = .732$, $\eta p^2 = .005$. Holm-adjusted pairwise comparisons (Figure 8) indicated greater calmness after the Calming than the Neutral condition, $t(55) = 2.88$, $p = .006$, $d = 0.38$, and markedly greater calmness after Calming than Stressful, $t(55) = 6.14$, $p < .001$, $d = 0.82$. Calmness was also higher in Neutral than Stressful, $t(55) = 4.71$, $p < .001$, $d = 0.63$.

As a robustness check, the linear mixed-effects model showed the same pattern, with a strong valence effect, $\chi^2(2) = 54.48$, $p < .001$, and no main effect of PSS, $\chi^2(1) = 0.48$, $p = .487$, and no Valence × PSS interaction, $\chi^2(2) = 0.55$, $p = .761$. Overall, these findings again support H1: emotional valence strongly modulated calmness, whereas H2 and H3 were not supported, as baseline stress (PSS) neither shifted calmness overall nor moderated valence effects.

**Effects on Energy**

Energy (i.e., arousal) ratings differed modestly by emotional valence in the repeated-measures ANCOVA, $F(1.88, 103.52) = 4.44$, $p = .016$, $\eta p^2 = .075$. The covariate effect was not significant, $F(1, 55) = 1.93$, $p = .170$, $\eta p^2 = .034$, and neither was the Valence × PSS interaction, $F(1.88, 103.52) = 0.71$, $p = .485$, $\eta p^2 = .013$. Holm-adjusted pairwise comparisons (Figure 8) indicated no difference between Calming and Neutral, $t(55) = -0.36$, $p = .723$, $d = -0.05$, but lower energy in Calming versus Stressful, $t(55) = -2.44$, $p = .040$, $d = -0.33$, and in Neutral versus Stressful, $t(55) = -2.56$, $p = .040$, $d = -0.34$.

As a robustness check, the linear mixed-effects model converged on the same pattern, showing a significant valence effect, $\chi^2(2) = 9.20$, $p = .010$, with no main effect of PSS, $\chi^2(1) = 2.00$, $p = .157$, and no Valence × PSS interaction, $\chi^2(2) = 1.48$, $p = .478$; Holm-adjusted contrasts similarly indicated higher energy in the Stressful condition than in Calming ($p = .022$) and Neutral ($p = .036$), with no Calming–Neutral difference ($p = .740$). Overall, H1 was partially supported for energy/arousal (Stressful > Calming/Neutral), whereas H2 and H3 were not supported, as baseline stress neither shifted energy overall nor moderated valence effects.

**Effects on Pupil Size**

A robust effect of emotional valence emerged for pupil size in the repeated-measures ANCOVA, $F(1.81, 99.48) = 357.85$, $p < .001$, $\eta p^2 = .87$. Neither the covariate, $F(1, 55) = 0.04$, $p = .833$, $\eta p^2 = .001$, nor the Valence × PSS interaction was significant, $F(1.81, 99.48) = 0.45$, $p = .618$, $\eta p^2 = .008$. Holm-adjusted pairwise comparisons (Figure 9) showed larger pupil diameters in Calming than Neutral, $t(55) = 21.09$, $p < .001$, $d = 2.82$; smaller diameters in Calming than Stressful, $t(55) = -6.80$, $p < .001$, $d = -0.91$; and the smallest diameters in Neutral compared with Stressful, $t(55) = -22.27$, $p < .001$, $d = -2.99$. Together with the VAMS shifts across conditions, these pupillometric differences provide an objective corroboration that the VR scenes elicited distinct emotional/attentional states.

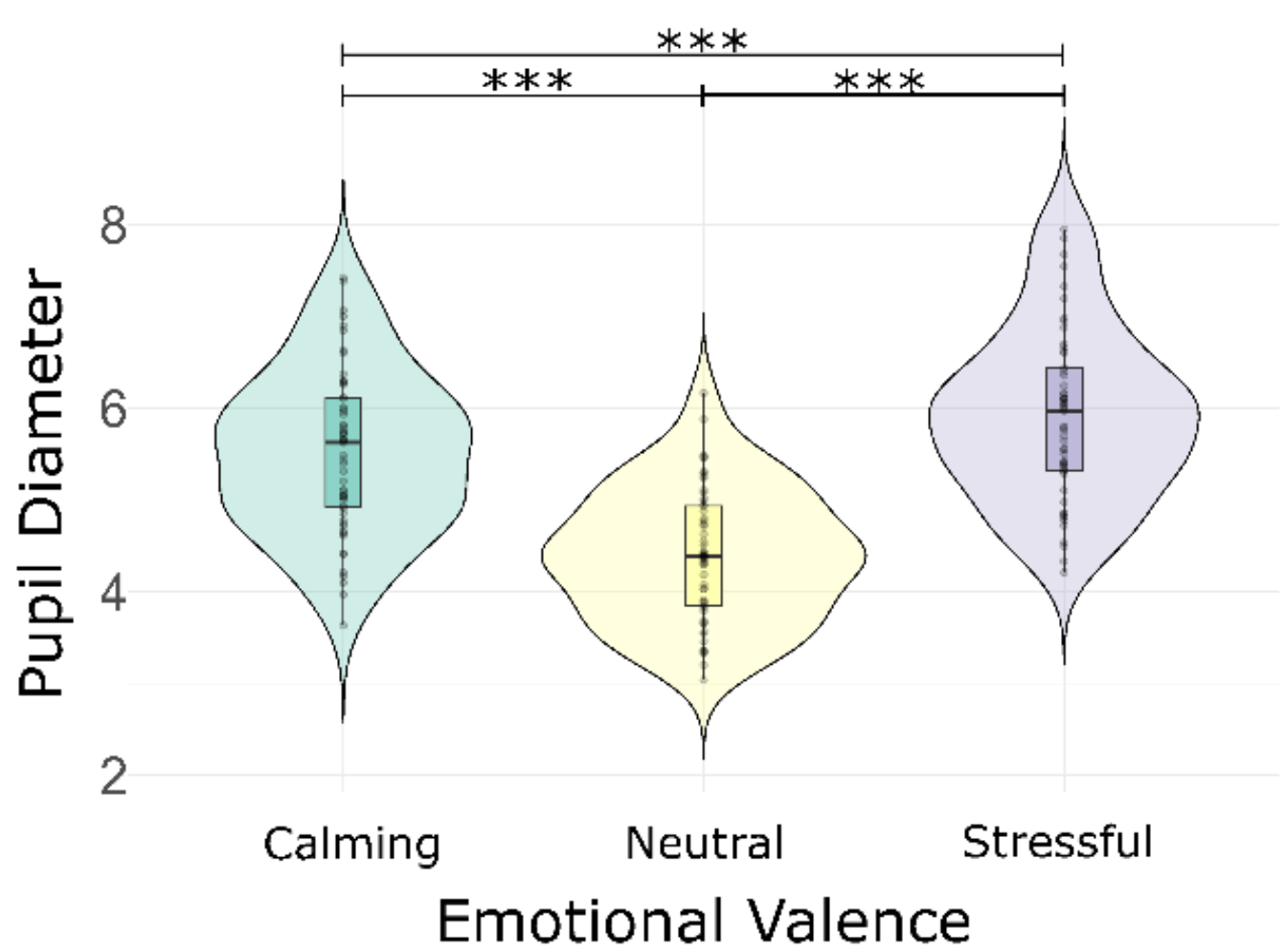

**Figure 9.** Emotional Valence Effects on Pupil Size/Diameter. *Measured in millimetres; *** p <.001.*

As a robustness check, the linear mixed-effects model replicated the same pattern, with a very large valence effect, $\chi^2(2) = 741.73$, $p < .001$, and no main effect of PSS, $\chi^2(1) = 0.05$, $p = .829$, and no Valence × PSS interaction, $\chi^2(2) = 0.94$, $p = .626$. Overall, these findings strongly support H1: physiological indices of arousal/attention varied systematically across emotional contexts, whereas H2 and H3 were not supported, as baseline stress neither shifted pupil size overall nor moderated valence effects.

**Effects on Time Passage Perception**

Time passage difference varied significantly with emotional valence in the repeated-measures ANCOVA, $F(1.78, 97.94) = 12.98$, $p < .001$, $\eta p^2 = .19$. Neither the covariate, $F(1, 55) = 0.03$, $p = .853$, $\eta p^2 < .001$, nor the Valence × PSS interaction was significant, $F(1.78, 97.94) = 0.50$, $p = .585$, $\eta p^2 = .009$. Holm-adjusted pairwise comparisons (Figure 10) showed greater overestimation in Calming than Neutral, $t(55) = 3.36$, $p = .003$, $d = 0.45$; greater overestimation in Calming than Stressful, $t(55) = 4.48$, $p < .001$, $d = 0.60$; and greater overestimation in Neutral than Stressful, $t(55) = 2.28$, $p = .026$, $d = 0.31$.

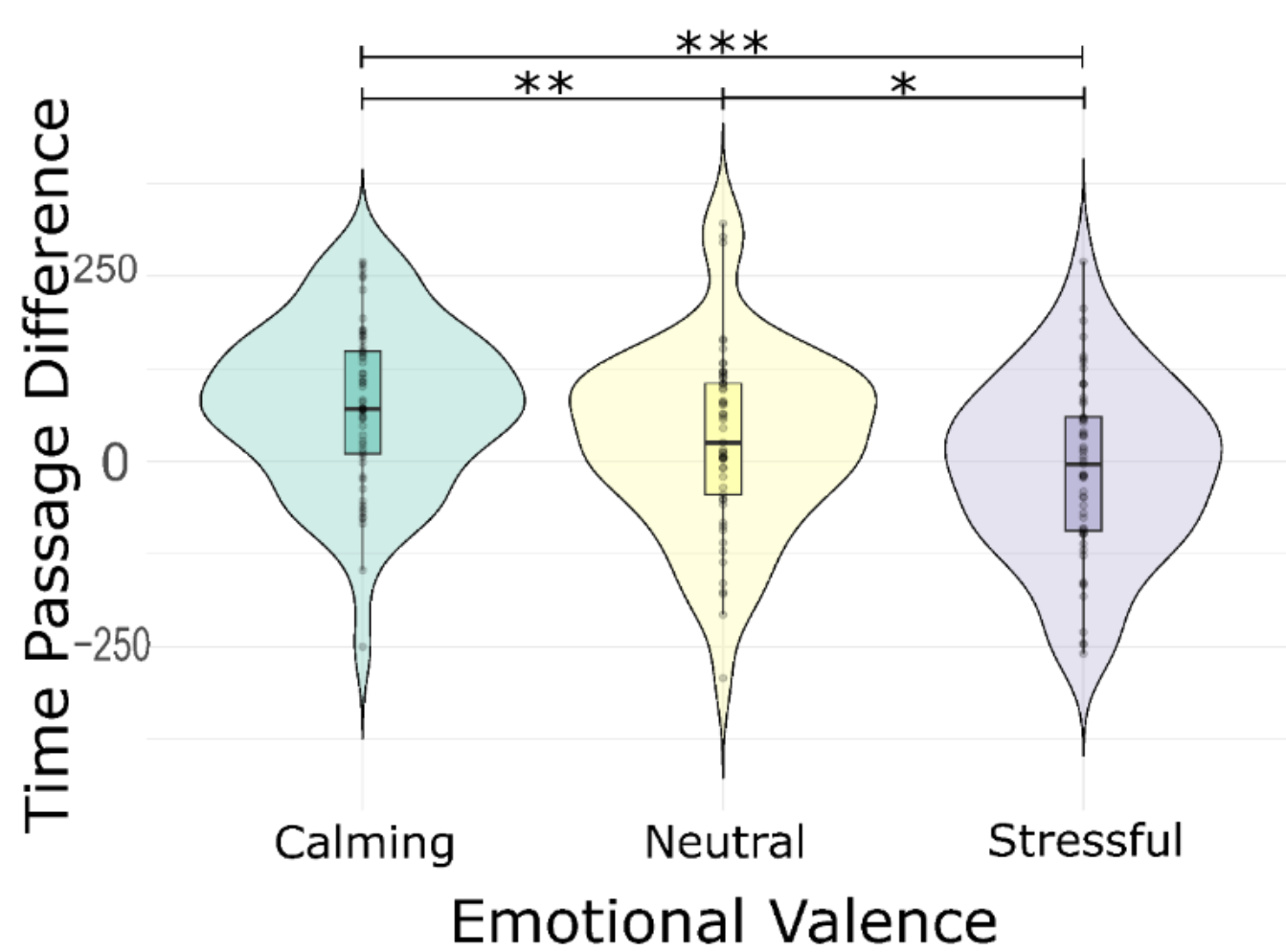

**Figure 10.** Emotional Valence Effects on Time Passage Perception. *Measured in seconds; * p <.05; ** p <.01; *** p <.001.*

As a robustness check, linear mixed-effects models with random intercepts for participants converged on the same pattern, showing a significant valence effect, $\chi^2(2) = 26.90$, $p < .001$, with no main effect of PSS, $\chi^2(1) = 0.04$, $p = .849$, and no Valence × PSS interaction, $\chi^2(2) = 1.04$, $p = .594$. Holm-adjusted contrasts were directionally identical; the Calming–Neutral difference remained significant but was slightly weaker in the LMM ($p$ = .013) than in the ANCOVA ($p$ = .003), while the other contrasts were consistent (Calming–Stressful, $p < .001$; Neutral–Stressful, $p \approx .02–.03$). Overall, these findings support H1: emotional valence systematically compressed or dilated subjective time, whereas H2 and H3 were not supported, as baseline stress did not predict time passage difference nor moderate valence effects.

## Discussion

This experiment compared three VR environments of contrasting emotional valence (calming garden, neutral room, stressful sewer) and tested whether baseline perceived stress predicted prospective time judgements and convergent affective/physiological responses within ICM and AGM frameworks. Emotional valence exerted a graded influence across all outcomes. Specifically, the calming garden produced the highest mood and calmness ratings and the greatest overestimation of elapsed time, whereas the stressful sewer yielded the lowest mood and calmness, elevated energy, the largest pupil dilations, and the strongest time compression, with the neutral room consistently intermediate. This pattern supports H1 and is consistent with the AGM, whereby emotionally salient contexts shift attentional allocation and thus gate width, altering the effective accumulation of timing information. In contrast, baseline stress showed no detectable main effects on time perception and did not moderate valence in this sample.

### Emotional Valence and Baseline Stress in Time Perception

Across outcomes, the environments produced a coherent valence-linked profile. Mood and calmness were highest in the Calming garden and lowest in the Stressful sewer, while pupil diameter robustly differentiated the contexts (Stressful > Calming > Neutral; [63]). Prospective time judgements showed the complementary behavioural pattern, with greater overestimation in the Calming context and greater compression in the Stressful context. Taken together, these convergent subjective, behavioural, and physiological shifts indicate that immersive emotional context reliably modulated prospective timing. Pupil diameter is best interpreted here as an objective marker of autonomic arousal [63] and attentional engagement [46], rather than as a direct proxy for perceived duration.

Within ICM/AGM accounts, the pattern is parsimoniously explained by attentional allocation. When a context strongly captures attention, the gate narrows and fewer pacemaker pulses reach the accumulator, biasing prospective judgements toward compression [10]. When more attentional capacity remains available for temporal monitoring, relative dilation becomes more likely [14, 15]. The Stressful scene was both highly arousing and plausibly the most attention-absorbing, because exploration required continuous tracking of the flashlight path and teleport nodes, and this aligns with the strong differentiation in pupil dilation.

This gate-dominant interpretation is consistent with evidence from high-engagement settings. Time-compression has been reported in other high-presence VR experiences, where immersive engagement plausibly reduces time monitoring even under arousing content [68, 79]. Findings from applied workload contexts point in the same direction. For example, arousal manipulations bias timing most clearly when paired with heavy monitoring demands (e.g., pilot a cockpit), indicating that attentional capture can be the decisive constraint on prospective estimates [69].

In contrast, baseline perceived stress showed no detectable main effects and did not moderate valence-related changes in the present sample. Our results suggest that any moderation—if present—would likely be small under these conditions. This pattern should not be read as evidence that baseline stress is irrelevant to prospective timing. Instead, tonic individual differences may be less likely to express when sustained immersive engagement dominates attentional resources across multi-minute exploration [42]. This boundary interpretation also fits accounts in which stress-related timing differences are more likely to emerge when attentional demands are lower or when intervals allow self-focused thought to compete with task engagement [106, 108]. In addition, restricted baseline-stress variability in healthy samples can further limit sensitivity to subtle moderation effects, particularly without complementary tonic biomarkers [78].

Finally, the results are compatible with motivational-intensity accounts in which stronger motivational pull curtails temporal monitoring [33], helping explain why the Stressful scene showed the greatest compression while the Calming scene showed relative dilation. Future work should test this more directly by orthogonally manipulating valence and motivational intensity and by quantifying VR-relevant scene properties (e.g., spatial openness/complexity, luminance/contrast), then fitting formal attentional-gate models to converging indices such as pupillometry and behavioural timing [112]. Such designs can also examine whether the present pattern generalises to longer intervals, where memory-based cues can become more influential [2, 21].

Taken together, the central inference is that, in immersive multi-minute VR exposures, attentional capture provides the most plausible proximate route through which emotional context shapes prospective time judgements. This conclusion motivates the next step, namely treating presence and agency as gain factors that may scale attentional capture and shift the balance between gate and pacemaker contributions in VR.

**An ICM × AGM × Presence Model for Time Perception in VR**

Immersive VR offers methodological advantages for timing research by enabling stronger affect induction and tighter control over engagement than many conventional paradigms. Modern head-mounted displays can elicit strong presence through sensorimotor contingencies and multi-sensory fidelity, supporting more naturalistic affective engagement while retaining experimental control [48, 88]. In this experiment, we observed graded gate-width shifts under controlled conditions and confirming that immersive engagement is sufficient to override modest trait differences in perceived stress.

These findings are consistent with the possibility that presence and agency operate as gain factors within the ICM × AGM architecture, scaling how strongly scenes capture attention. Presence—the lived sense of “being there” grounded in sensorimotor contingencies [86]— can amplify attentional capture and emotional salience, thereby shaping gate dynamics in attentional timing accounts. In our scenes, the Stressful environment showed higher arousal/engagement indices (energy ratings and pupil dilation) together with stronger time compression, whereas the Calming environment showed the opposite behavioural tendency—patterns aligned with the view that attentional capture can dominate arousal-linked pacemaker influences during immersive engagement and determine the direction of the distortion [35, 54, 68, 79].

A complementary implication is that when presence or agency is degraded, attention is less monopolised and arousal-linked pacemaker acceleration may become more visible in timing judgements. This boundary condition is consistent with evidence from less immersive paradigms, where reduced engagement and constrained interaction can shift the balance toward pacemaker-driven effects on perceived duration [64]. Taken together, these bidirectional patterns suggest that presence/agency modulates the *relative weight* of pacemaker speed and gate width, rather than acting as a simple additive influence.

Building on ICM and AGM accounts, we therefore propose—tentatively—a three-factor ICM–AGM–Presence model for timing distortions in immersive settings. In this framework, arousal governs pacemaker rate, attention governs gate width, and presence/agency amplifies—or attenuates—the extent to which external events capture attention. When presence is high, motivational intensity may help calibrate how valence translates into timing via attentional capture. Presence thus does not cancel valence; it amplifies the attentional pull of a scene's motivational intensity—compressing time in high-arousal contexts and dilating it in low-arousal ones, with affectively neutral settings anchored between.

Future studies can manipulate presence directly—for example by reducing graphics resolution, altering avatar embodiment, or adding controller latency—while varying agency via self-initiated versus guided locomotion. Because high-presence VR can approximate naturalistic sensorimotor contingencies while preserving laboratory control, it provides a promising bridge between mechanistic timing theory and applied immersive design [48, 80]. The approach therefore promises an uncommon blend of ecological validity, mechanistic precision, and practical impact.

**Practical Implications for XR Design, Digital-Wellbeing, and "Time-Loss" Safeguards**

The valence–timing profile observed here provides a concrete dial for XR creators. Low-arousal, positive-valence scenes (e.g., natural vistas, slow music, pastel palettes) shifted subjective time, whereas high-arousal, negative-valence scenes (e.g., narrow corridors, suspenseful audio) produced the strongest time compression. This implies that colour, soundscape, and motion cues can be tuned not only for emotional tone but also for perceived session length, consistent with VR distraction/analgesia work showing that immersive attentional capture can reduce aversive waiting and procedural burden [84]. For advertising, this offers a route to tune perceived dwell time and narrative pacing, potentially increasing message exposure without extending objective duration. For professional training (e.g., safety, emergency response), designers can calibrate scenario intensity to avoid unintended time compression that could distort perceived workload and debrief accuracy.

The same data suggest that pupil dilation can serve as a built-in telemetry channel for adaptive experience design. Because pupil diameter robustly differentiated the emotional contexts, pupillometry in eye-tracking–enabled headsets can be used as a continuous index of autonomic arousal and attentional engagement. A simple pipeline could reference each user to a neutral baseline and intervene when dilation remains elevated (e.g., schedule a micro-break in learning, soften difficulty in rehabilitation, or smooth reward spikes in entertainment). Evidence that pupillometry-based biofeedback can reduce anxiety supports the feasibility of embedding such regulation loops inside XR experiences [60].

More broadly, pronounced "time loss" is frequently discussed in models of problematic technology use, and immersive systems may intensify attentional capture in some contexts [12, 48]. A feasible safeguard could be to combine objective elapsed time with sustained elevations in pupil-based engagement to trigger unobtrusive pacing interventions (e.g., brief pauses, content transitions), aligning with digital-wellbeing approaches [48]. Longitudinal studies pairing eye-tracking with usage logs and craving/urge indices can calibrate thresholds that precede problematic use, and related mobile-nudge work indicates technical feasibility [41, 71, 76].

**Limitations and Future Directions**

Our sample comprised healthy young adults, limiting generalisability to older cohorts and clinical populations. In addition, baseline perceived stress often shows restricted variance in non-clinical student samples, which can reduce sensitivity to small main effects or moderation by stress. Future studies should recruit cohorts with broader baseline-stress variability and add tonic biomarkers (e.g., heart-rate variability and/or cortisol) to better characterise stress-related individual differences.

Exposure duration was fixed at five minutes and locomotion relied on teleportation. This supported experimental control and reduced motion-related confounds, but it constrains inference to other interval lengths and more naturalistic movement. Future work should systematically vary exposure duration, contrast teleportation with continuous locomotion, and include presence/ecological realism and plausibility/agency ratings to quantify perceived naturalism under different interaction constraints.

Prospective timing was assessed with a single verbal estimation task, and the environments necessarily differed in geometry and audiovisual structure to elicit distinct affective contexts. Future work should replicate the pattern across multiple prospective timing paradigms (e.g., reproduction or bisection), collect brief ratings of novelty, complexity, threat, and presence/agency, and quantify core scene statistics (e.g., luminance/contrast and motion energy) to better separate affective influences from structural variation. Finally, adding complementary autonomic and behavioural measures (e.g., GSR/EDA, heart rate, startle, approach–avoidance behaviour) alongside pupillometry would strengthen mechanistic inference.

## Conclusions

Immersive VR provided an ecologically valid and controlled testbed for timing theory. Across scenes of diverse emotional valence, a monotonic pattern emerged: the calming garden improved mood and lengthened perceived duration, the neutral room was intermediate, and the stressful sewer increased arousal and produced the strongest time compression. This refines the integrated Internal Clock × Attentional Gate account by indicating that, under strong presence and agency, prospective timing is governed primarily by moment-to-moment attentional capture rather than by pacemaker speed or baseline perceived stress. Emotional valence in immersive VR can therefore modulate attentional allocation in ways that shape prospective timing, while baseline perceived stress did not show evidence of meaningful influence in this sample.

From an applied perspective, emotional valence may modulate subjective time perception within XR applications—ranging from training and rehabilitation to advertising—where perceived session length, engagement, and cognitive absorption can be intentionally shaped. Moreover, pronounced distortions in perceived time may serve as an implicit marker of deep engagement, a design parameter, and a safety checkpoint in the development of next-generation XR applications. Finally, this study's findings propose an ICM × AGM × Presence Model, which situates time perception at the intersection of internal-clock mechanisms, attentional gating, and immersive presence, and provides a foundation for future work on how emotion, attention, and embodiment jointly shape the perceived passage of time in immersive VR.

## Acknowledgements

The authors thank the participants for generously offering their time, as well as the student research assistants of the PsyNet Lab for their help with data collection.

## Competing interests

The authors declare that they have no financial or non-financial competing interests.

## Financial support

No external grant or industry sponsorship was received.

## CRediT author statement

**Kyriaki Syrigou:** Conceptualization; Investigation; Data curation; Formal analysis; Writing – Original Draft; Writing – Review & Editing.

**Marina Stoforou:** Investigation; Data curation; Formal analysis; Writing – Original Draft; Writing – Review & Editing.

**Panagiotis Kourtesis:** Conceptualization; Methodology; Software; Formal analysis; Visualization; Writing – Original Draft; Writing – Review & Editing; Supervision; Project administration.